\documentclass[10pt,aps,pra,twocolumn,showpacs,amsmath,amssymb,footinbib]{revtex4-1}
\usepackage{color}
\usepackage[usenames,dvipsnames]{xcolor}
\usepackage{graphicx}
\usepackage{epstopdf,mathrsfs}
\usepackage{booktabs}
\usepackage{bbold}
\usepackage{subfig}
\usepackage{dcolumn}
\usepackage{bm}
\usepackage{braket}
\usepackage{CJK}
\usepackage{amsmath,amsfonts,amssymb,amsthm}
\usepackage{mathptmx}
\usepackage{calc}
\usepackage{tikz}
\usepackage{hyperref}
\usepackage{dsfont}
\usepackage{amssymb}
\usetikzlibrary{tikzmark,calc,,arrows,shapes,decorations.pathreplacing}
\tikzset{every picture/.style={remember picture}}
\usepackage{setspace}
\begin{document}
\title{Quantum coherence of spatial photonic qudits: experimental measurement and path-marker analysis}
\author{P. Machado}
\author{S. P\'adua}
\affiliation{\textsuperscript{}Departamento de F\'isica, Universidade Federal de Minas Gerais, 31270-901 Belo Horizonte, Minas Gerais, Brazil \date{\today}
}

\begin{abstract} 

Discussions about quantum interference, indistinguishability and superpostion between quantum states goes back to the beginning of quantum mechanics, but the theoretical problem concerning quantitative measures for quantum coherence was only recently solved by Baumgratz \textit{et al.} [\href{https://doi.org/10.1103/PhysRevLett.113.140401}{Phys. Rev. Lett. \textbf{113} 140401 (2014)}]. Since then many works have explored one of the possible coherence measures, the $l_1$ norm, which has not yet been experimentally obtained for spatial photonic states in high dimensions. In this article we study states prepared with photons crossing multiple slits and implement the theoretical proposal of T. Paul and T. Qureshi [ \href{https://doi.org/10.1103/PhysRevA.95.042110}{Phys. Rev. A \textbf{95} 042110 (2017)}] to determine experimentally the $l_1$ norm of qutrit states. We analyze the method validity and present an alternative for states in which one of their assumptions does not apply. We also discuss the requirements necessary for using an interference pattern for measuring quantum coherence of qudits by treating the measure-operator that describe the detection optical configuration as path-markers, relating them with the quantum coherence determination.

\end{abstract}

\maketitle

\section{Introduction}\label{introdu}
Quantum coherence is a feature of quantum systems related to an inability to distinguish between two states of a quantum superposition with two or more states \cite{1991Mandel}. Historically, it has been associated with the indistinguishability of quantum trajectories in interferometers. Primarily motivated by questions concerning particle-wave dual behavior, interferometric dualities have been a subject of study since the beginning of quantum mechanics \cite{1979Wooters,1980Bartell}.

In the $90$'s it was well established that for two-dimensional states, or interferometers with two trajectories, the visibility of the observed interference pattern quantifies the indistinguishability between the two states in superposition, that is, between the two trajectories in the interferometer \cite{1988Yasin}. Englert was the first to show quantitatively that the observation of interference fringes depends, not only on the state studied, but also on the states of the detector used for the measurements \cite{1996Englert}. Despite quantum coherence being a state property of a quantum system, the observed quantum coherence also depends on the measurement apparatus and may be less than or equal to that belonging to the superposition state.

Over the years, many authors tried without success to generalize Englert quantifiers to higher dimension quantum states \cite{2001-Durr, 2003-BimontePRA, 2003-Bimonte}, but only in $2014$ Baumgratz, Cramer and Plenio obtained adequate measures for quantum coherence through resource theory \cite{2014Baumgratz, 2015-Querish}. Although the theoretical problem has been solved, the question about the experimental measurement of quantum coherence of higher dimension systems remains.   
 
One of the quantum coherence measures presented in Ref.\cite{2014Baumgratz} was demonstrated experimentally in Ref.\cite{2017Guo}, in a context of photonic states in polarization variables. Concerning transverse spatial photonic states, a theoretical method for measuring quantum coherence was proposed in Ref.\cite{2017TaniaPaul}. In this article we implement this method for qutrit states and analyze its validity, presenting an alternative for states in which the proposal of Paul and Qureshi is not applicable. 

The article is organized as follows: 
In section \ref{methods} we analyze the link between quantum coherence and the interference pattern in detail. 
In section \ref{experimento} we present, for some qutrit states, results obtained experimentally based on the proposal of Paul and Qureshi. In sequence, in section \ref{alternative}, we discuss an alternative way to measure quantum coherence for cases in which this method is not applicable. Finally, we conclude in section \ref{conclusion}. In the Appendix \ref{apendice}, we show that we can regard the system passing through the measurement apparatus, as an interaction between the photonic path state and a path-marker and what are the requirements for the measurement apparatus such that the interference pattern is observed and quantum coherence is measured.
  
\section{Quantum coherence and photonic interference pattern}\label{methods} 
\begin{figure}[h!]
\centering
\includegraphics[height=12cm]{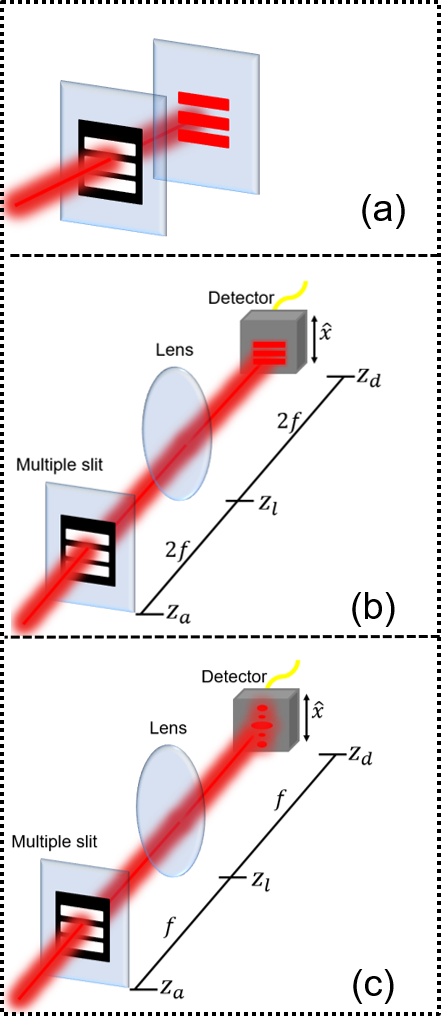}
\caption[Esquema de preparação de estado de fenda]{
{\footnotesize 
a) Schematic representation of photons linear momentum discretization by a multiple slit set. Photons pass through multiple slits positioned transversely with respect to the direction of photons propagation. Only photons with linear momentum within a specific range of values pass through some aperture, and $\ket{l}$ is the state of a photon crossing a specific aperture $l$. b) A lens with a focal lengh $f$ is placed at distance $f/2$ from both the multiple slit and the detector plane alowing measurement of the photonic spatial probability distribution at the image plane. c) A lens with a focal lengh $f$ is placed at distance $f$ from both the multiple slit and the detector plane alowing measurement of the photonic spatial probability distribution at the Fourier plane, i. e., multiple slit the interference pattern.}}\label{fig:discrete}
\end{figure}

Quantum Coherence is a property of a quantum state system. Its measurement depends on the observation of the interference pattern between  the vectors states components (“paths”) of the system. However, the interference pattern observation requires a path-marker being in a superposition state such that the detection system is unable to distinguish the state component in which the system is (or which interferometer path the particle followed), as pointed out by Englert \cite{1996Englert}. In this article, we deal with photons in a state superposition of transverse path states. The state system is prepared after a photon crosses a multiple slit in such a way that the state components are defined by the possible photon paths through the apertures of the multiple slit set (Fig. \ref{fig:discrete}.a). The quantum coherence measurement is done by measuring interference patterns using a lens placed in the path of the photon after it crosses the multiple slit towards the detection plane (Fig. \ref{fig:discrete}.c) . The ratio between the distance multiple slit set-lens plane and the distance lens plane-detector plane is essential (Figs. \ref{fig:discrete}.b and \ref{fig:discrete}.c). In the Appendix, we demonstrate that the propagation of the spatial photon state through the free space after crossing a lens in direction of the detector plane plays a role as a path-marker. We demonstrate that it is possible to observe the state quantum coherence only in the condition that the possible path-marker states are maximally indistinguishable. This condition occurs when the detector is at the Fourier plane, i. e., at the focal plane of a lens or at the far field plane (Fig. \ref{fig:discrete}.c). The possibility to infer quantum coherence from an interference pattern, where the possible path-marker states are non-orthogonal and maximally indistiguishable, accompanies the question: how to determine the value of the quantum coherence from the detected spatial interference pattern, i. e., from the photonic spatial probability distribution at the Fourier Plane? We aim to answer this question in this section.

A simple method to prepare spatial photonic states consists into discretizing the transverse profile of an ensemble of photons, identically generated, using multiple slits \cite{2005leo5, 2007leo2, 2008Japoneses, holandeses}. Fig. \ref{fig:discrete}.a shows a multiple slits array which selects photons within specific intervals of linear momentum. We can describe the state immediately after the apertures by the density operator
\begin{eqnarray}\label{estadogeral}
\rho &=& \sum_{l=-\lambda}^{\lambda}\sum_{m=-\lambda}^{\lambda}\rho_{lm}\ket{l}\bra{m},
\end{eqnarray}
where $\lambda=\frac{D-1}{2}$, $D$ is the dimension of the state given by the number of slits, $\left\{\rho_{lm}\right\}$ is the set of the density matrix coefficients and $\left\{\ket{j}\right\}$ is the chosen basis with $\ket{j}$ indicating here the state of a photon transmitted through the aperture $j$. We use a multiple-slit set with thin apertures with $2a$ width, so we can consider only the detection of photons that crossed the multiple-slit set with $D$ apertures. After photons are transmitted through slits their path states forms a discrete state set. 
In this scenario, the probability of detecting one photon on the detection plane at the longitudinal position $z_d$ and along the transverse $x$-direction is \cite{2000sebastiao1,1995Mandel,leo04,leo20072}
\begin{eqnarray}\label{Pgeral}
P\left(x,z_d\right)&=&Tr\left(\Gamma\rho\right),  
\end{eqnarray}
where $\Gamma=E^{\left(\scriptstyle{-}\right)}\left({\scriptstyle x,z_d}\right)E^{\left({\scriptstyle +}\right)}\left({\scriptstyle x,z_d}\right)$ is a positive operator describing the measurement
and  $E^{(+)}(x,z_d)$ is the electrical field operator proportional to the annihilation operator \textit{a}($x$) that indicates the destruction of a photon at the transverse position $x$ in the plane at $z_d$, which is the plane where the detector is scanned along $x$-direction. 
The density operator $\rho$ is written in terms of the $\ket{j}$ defined at the multiple-slit plane and $E^{(+)}(x,z_d)$ is constructed from the multiple slit to the detection plane and therefore includes the optical configuration that guides photons from the multiple slit to the detector. Fig. \ref{fig:discrete}.b and \ref{fig:discrete}.c show respectively two possible optical configurations: detection of the photons at the image plane and at the Fourier Plane.

The probability ditribution of detecting a photon at $x$ transverse position when the detector is scanned at the Focal plane\cite{leo20072,miguel2010},

\begin{eqnarray}\label{sec2padrao}
P(x,z_D)
&\propto&\textit{sinc}^2\left(\frac{kax}{f}\right)\left[\sum_{m}\rho_{mm} +\right.\nonumber\\
&&\left.+\sum_{l\neq m,l<m}2\left|\rho_{lm}\right|\cos\left(\gamma \left(l-m\right)x-\varphi_{lm}\right)\right],
\end{eqnarray}
where $\sum_{m}\rho_{mm}=1$, $\varphi_{lm}$ is the phase of $\rho_{lm}$, $k$ is the wavenumber of the photons, $f$ is the focal length of the lens, and $\gamma$ is the phase that the photons acquire propagating from the slits to the plane at $z_d$ \cite{2010Leo}.

Quantum coherence defined in Ref. \cite{2014Baumgratz} for the state represented in Eq.(\ref{estadogeral}), in a normalized form \cite{2015-Quresh}, is
\begin{eqnarray}\label{coherence}
\mathcal{C} &=& \frac{1}{D-1}\sum_{l=-\lambda}^{\lambda}\sum_{m\neq l=-\lambda 
}^{\lambda}\left|\rho_{lm}\right|.
\end{eqnarray}

In order to extract the quantum coherence $\mathcal{C}$ from Eq. \ref{sec2padrao}, we define an oscillation function $\Theta\left(x,z_d\right)$, namely
\begin{eqnarray}\label{oscfuntion}
\Theta\left(x,z_d\right)&=&\frac{P(x,z_d)-P_{diag}(x,z_d)}{P_{diag}(x,z_d)}\nonumber\\
&=&\sum_{l<m} 2\cos\left(\gamma\left(l-m\right)x-\varphi_{lm}\right)\left|\rho_{lm}\right|,
\end{eqnarray}
where $P_{diag}(x,z_d)$ is the probability to detect one photon from a state $\rho_{diag}$ on the far-field plane (Eq. (\ref{sec2padrao})). The state $\rho_{diag}$ is an auxiliary state, supposed to be previously characterized as incoherent and having the same diagonal elements of $\rho$. We consider that both $\rho$ and $\rho_{diag}$ are normalized in Eq. (\ref{oscfuntion}).

From now on we will analyze a qutrit photon state in path variables, as an example, but one could extend our discussions for higher dimensions without difficulties. For qutrit systems, we can write a general and normalized state by
\begin{equation}\label{pho3}
\rho=\bordermatrix{~ & \bra{-1} & \bra{0} & \bra{1}\cr
\ket{-1}&\rho_{{\scriptscriptstyle-1-1}} & \left|\rho_{{{\scriptscriptstyle -10}}}\right|e^{i\varphi_{{\scriptscriptstyle-10}}} & \left|\rho_{{\scriptscriptstyle-11}}\right|e^{i\varphi_{{\scriptscriptstyle-11}}}\cr
\ket{0}&\left|\rho_{{\scriptscriptstyle-10}}\right|e^{-i\varphi_{{\scriptscriptstyle-10}}} & \rho_{{\scriptscriptstyle 00}} & \left|\rho_{{\scriptscriptstyle 01}}\right|e^{i\varphi_{{\scriptscriptstyle 01}}}\cr
\ket{1}&\left|\rho_{{\scriptscriptstyle -11}}\right|e^{-i\varphi_{{\scriptscriptstyle -11}}} & \left|\rho_{{\scriptscriptstyle 01}}\right|e^{-i\varphi_{{\scriptscriptstyle 01}}} & \left|\rho_{{\scriptscriptstyle11}}\right|},
\end{equation}
and its oscillation function as  
\begin{eqnarray}\label{osc3}
\Theta\left(x,z_d\right)&=&2\left|\rho_{{{\scriptscriptstyle -10}}}\right|\cos\left(\gamma x+\varphi_{{\scriptscriptstyle-10}}\right)+2\left|\rho_{{{\scriptscriptstyle 01}}}\right|\cos\left(\gamma x+\varphi_{{\scriptscriptstyle 01}}\right)\nonumber\\
&&+2\left|\rho_{{{\scriptscriptstyle -11}}}\right|\cos\left(2\gamma x+\varphi_{{\scriptscriptstyle -11}}\right),
\end{eqnarray}

\begin{figure}[h!]
\centering
\includegraphics[height=8.5cm]{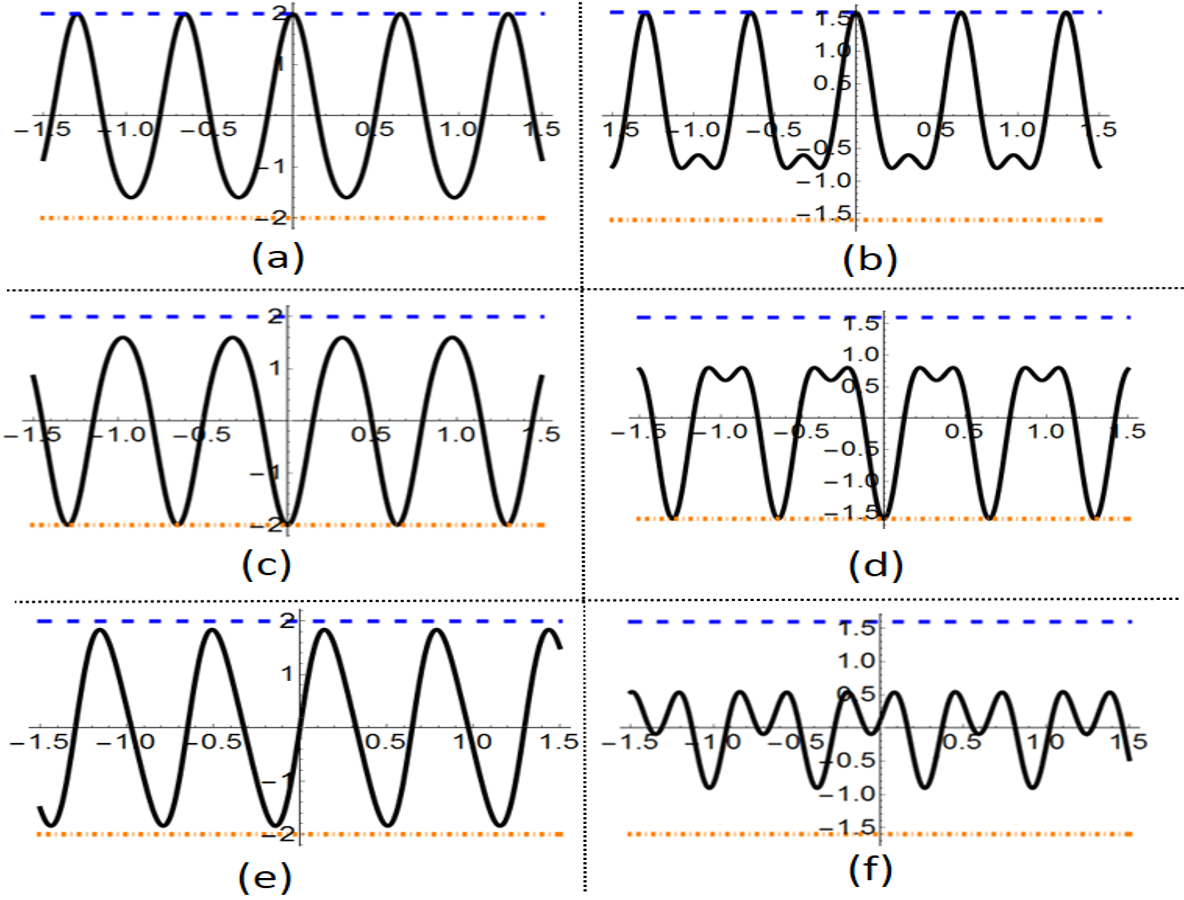}  
\caption[Oscilações teóricas para um qutrit]{{\footnotesize  Theoretical curves of the oscillation function $\Theta\left(x,z_D\right)$ in Eq.(\ref{osc3}) for qutrit states. The blue dashed lines (on top of the curves) mark the amount of the qutrits quantum coherence (in a non normalization form) and the dot dashed orange lines mark the negative of this value. In (a) and (b) all phases $\varphi_{lm}$ are equal to zero,    
 \textit{i.e.}, $\left\{\varphi_{lm}=0\right\}$. In (c) and (d) $\left\{\varphi_{lm}=\pi\right\}$,  in (e) $\left\{\varphi_{lm}=\pi/2\right\}$, and in  
(f) $\varphi_{-11}=2\pi/3,\varphi_{-10}=\pi/2,\varphi_{01}=5\pi/3$. In (b), (d) and (f), $\left|\rho_{{{\scriptscriptstyle -11}}}\right|=0.25$, $\left|\rho_{{{\scriptscriptstyle -10}}}\right|=0.2$, $\left|\rho_{{{\scriptscriptstyle 01}}}\right|=0.35$, and in (a), (c) and (e), $\left|\rho_{{{\scriptscriptstyle -11}}}\right|=0.4$, $\left|\rho_{{{\scriptscriptstyle -10}}}\right|=0.1$, $\left|\rho_{{{\scriptscriptstyle 01}}}\right|=0.5$.}}\label{fig:oscmatematica}
\end{figure}

Fig.\ref{fig:oscmatematica} shows the qutrit oscillation function (Eq. (\ref{osc3})) for different states, that is, qutrit states with different coefficients $\rho_{lm}$ and phases $\varphi_{lm}$. In Figs. \ref{fig:oscmatematica}(a) and (b) all the phases $\varphi_{lm}$ are equal to zero, \textit{i.e.}, $\left\{\varphi_{lm}=0\right\}$. In Figs. \ref{fig:oscmatematica}(c) and (d) $\left\{\varphi_{lm}=\pi\right\}$, in Fig. \ref{fig:oscmatematica}(e) $\left\{\varphi_{lm}=\pi/2\right\}$, and in 
Fig. \ref{fig:oscmatematica}(f), $\varphi_{-11}=2\pi/3,\varphi_{-10}=\pi/2,\varphi_{01}=5\pi/3$. In Figs. \ref{fig:oscmatematica}(b), (d) and (f) $\left|\rho_{{{\scriptscriptstyle -11}}}\right|=0.25$, $\left|\rho_{{{\scriptscriptstyle -10}}}\right|=0.2$, $\left|\rho_{{{\scriptscriptstyle 01}}}\right|=0.35$, 
and in Figs. \ref{fig:oscmatematica}(a), (c) and (e) $\left|\rho_{{{\scriptscriptstyle -11}}}\right|=0.4$, $\left|\rho_{{{\scriptscriptstyle -10}}}\right|=0.1$, $\left|\rho_{{{\scriptscriptstyle 01}}}\right|=0.5$. In all the graphs, the top lines (dashed blues lines) correspond to the respective quantum coherence in a non normalized form, namely, the value of $2\left(\left|\rho_{{{\scriptscriptstyle -11}}}\right|+\left|\rho_{{{\scriptscriptstyle -10}}}\right|+\left|\rho_{{{\scriptscriptstyle 01}}}\right|\right)$. The bottom lines (dot dashed orange lines) correspond to the negative of these values. 

For states with null-phases in Figs. \ref{fig:oscmatematica}(a) and (b) (or phases equal to $\pi$ in Figs. \ref{fig:oscmatematica}(c) and (d)), we can determine quantum coherence by the maximum (or the minimum) value achieved by the oscillation curve. We can write

\begin{equation}\label{oscmaxemin}
\mathcal{C} = \begin{cases}
\frac{\Theta_{max}\left(x_{max},z_d\right)}{D-1} \ \mbox{{\small when}}\ \left\{\varphi_{lm}=0\right\},\\ \\
\frac{\Theta_{min}\left(x_{min},z_d\right)}{D-1} \ \mbox{{\small when}}\ \left\{\varphi_{lm}=\pi\right\}.
\end{cases}
\end{equation}

This is possible because in the case of phases of the density matrix elements equal to $0$ (or $\pi$) each cosine term in Eq. (\ref{osc3}), reaches a maximum (or minimum) value at the same point $x_{max}$ ($x_{min}$). It makes the overall maximum (or overall minimum) equal to the sum of the absolute values of the out-diagonal elements of $\rho$. 

The determination of $l_1$ norm quantum coherence by means of the central maximum of the interference pattern in Eq.(\ref{sec2padrao}) and the diffraction pattern from $\rho_{diag}$, was theoretically proposed in \cite{2017TaniaPaul}, that is, the authors proposed the use of what is equivalent to a single point of oscillation curve in Eq.(\ref{oscfuntion}). The authors treated the case where the phases of all density matrix elements are null. However, Fig.\ref{fig:oscmatematica} shows that if the state $\rho$ has phases different from zero (or $\pi$), or even if it is completely unknown, we cannot determine its quantum coherence by using the plot of its oscillation function, neither employing the contrast of the interference patterns.

In the next section we describe an experiment for $l_1$ norm quantum coherence determination, considering qutrit states for which Eq.(\ref{oscmaxemin}) is valid. In Sec. \ref{alternative}, we present an alternative for the cases for which Eq.(\ref{oscmaxemin}) is not applicable.

\section{ Experiment with qutrits}\label{experimento}

\begin{figure}[h!]
\centering
\includegraphics[height=6cm]{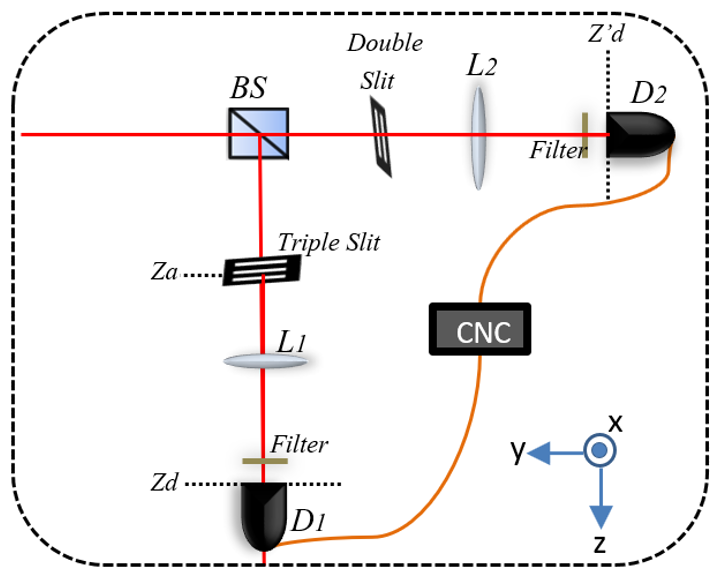}
\caption[Setup do experimento]{{\footnotesize Experimental setup for preparing and detecting qutrit states. A beam splitter (BS) separates twin photons generated by SPDC, in a collinear phase matching. $810$ nm interference filters with bandwidth of $10$ nm are used to select the photon pairs. Each photon of a pair goes to a different multiple-slit. In the plane of the multiple-slit at $z_a$, immediately after the apertures, different photonic spatial correlations are prepared as shown in Ref. \cite{2018Paula}. The photon in the qubit state passes trough a lens ($L_3$) that projects the slits image on a detection plane at $z'_d$. The photon in the qutrit state passes trough a lens ($L_1$) that realizes the optical Fourier transform on a detection plane at $z_d$. 100 $\mu$m single slit is coupled to the detector $D1$ such that the spatial fringes can be resolved. The qubit detector ($D_2$) is a bucket detector which collects all transmitted photons such that the counts of qubit photons work as a trigger to the counts of the qutrit photons. The qutrit detector ($D_1$) records the photons transverse distribution by scanning $D_1$ along the $x$-direction.}}\label{fig:setupcoe}
\end{figure}

We determined quantum coherence of two different spatial qutrit states which are, each of them, part of a bipartite $2\times 3$ state prepared with multiple slits and twin photons generated by Spontaneous Parametric Down Conversion (SPDC) that cross the slits \cite{2018Paula}. Photons are generated in a collinear regime by type I SPDC phase matching \cite{2009HPires}. Fig. \ref{fig:setupcoe} shows a beam splitter ($50:50$ BS) separating the photons of a pair in such a way that each photon can be transmitted or reflected by the BS with probability of almost $50 \%$. Photons transmitted through the BS go to the double slit and the reflected ones go to a triple slit. Different manipulations of the pump beam transverse profile and of the propagated phase matching function provide different photon path states behind the slits \cite{2018Paula}.

From $2\times 3$ states, we can obtain spatial qutrit states by tracing out the spatial qubit degrees of freedom. Considering the same base state of Eq. (\ref{pho3}), the qutrit density matrix that we obtained marginally from the $2\times 3$ states prepared as in Ref.\cite{2018Paula}, have the following theoretical forms
\begin{equation}\label{states}
\rho_{I}=\begin{pmatrix} \rho_{{\scriptscriptstyle-1-1}} & \left|\rho_{{{\scriptscriptstyle -10}}}\right| & 0 \\ 
\left|\rho_{{\scriptscriptstyle-10}}\right| & \rho_{{\scriptscriptstyle 00}} & \left|\rho_{{{\scriptscriptstyle 01}}}\right| \\ 
0 & \left|\rho_{{{\scriptscriptstyle 01}}}\right| & \rho_{{{\scriptscriptstyle 11}}}\end{pmatrix}
,\ 
\rho_{II}=\begin{pmatrix} \rho_{{\scriptscriptstyle-1-1}} & 0 & \left|\rho_{{\scriptscriptstyle -11}}\right| \\ 
0 & \rho_{{\scriptscriptstyle 00}} & 0 \\ 
\left|\rho_{{\scriptscriptstyle -11}}\right| & 0 &\rho_{{\scriptscriptstyle 11}}\end{pmatrix},
\end{equation}
which have specific symmetries that lead, respectively, to the following oscillation functions 
\begin{eqnarray}\label{osccadaum}
\Theta_{I}\left(x,z_D\right)&=&\mathcal{A}_1\cos\left(\gamma x\right),\nonumber\\
\Theta_{II}\left(x,z_D\right)&=&\mathcal{A}_2\cos\left(2\gamma x\right),
\end{eqnarray}
where $\mathcal{A}_1=2\left(\left|\rho_{{{\scriptscriptstyle -10}}}\right|+\left|\rho_{{{\scriptscriptstyle 01}}}\right|\right)$ and $\mathcal{A}_2=2\left|\rho_{{{\scriptscriptstyle -11}}}\right|$ are the respective oscillation amplitudes.

The states $\rho_{I}$ and $\rho_{II}$ produce oscillations ($\Theta_{I}$ and $\Theta_{II}$) equivalent to two-dimensional spatial states, even though their Hilbert space dimension is equal to three. This happens because the interference pattern of $\rho_{I}$ has all cosine terms with the same spatial frequency since the off-diagonal elements $\left|\rho_{{{\scriptscriptstyle -11}}}\right|$ and $\left|\rho_{{{\scriptscriptstyle 1-1}}}\right|$ are null. This qutrit state is prepared by having $\ket{-1}$ and $\ket{1}$ distinguishable. The interference pattern produced by $\rho_{II}$ has only one cosine term since now the only off-diagonal elements non null are $\left|\rho_{{{\scriptscriptstyle -11}}}\right|$ and $\left|\rho_{{{\scriptscriptstyle 1-1}}}\right|$. In this case, we prepare this qutrit state by having only the states $\ket{-1}$ and $\ket{1}$ indistinguishable.

Because the states in Eq. (\ref{states}) have null phases, it is possible to determine their quantum coherence by means of the functions in Eq.(\ref{osccadaum}). Further, their symmetries also enable us to determine it employing the usual contrast of a typical two-dimension interference patterns \cite{livrofowles}, which we use for comparing the results from the different methods used for measuring quantum coherence and to evaluate the reliability of the method that we present in this work. Therefore, we study qutrit states described by simple density matrices, and with simple interference patterns, aiming to verify if  the method works well and if it could be applied to any qudit state with the matrix element phases equal to zero or $\pi$.

Fig. \ref{fig:setupcoe} shows our experimental setup for quantum coherence determination. A lens $L_2$ projects the double slit image on the detection plane at $z'_{d}$, where an avalanche photodiode detector ($D_2$) is maintained fixed. This detector acts as a bucket detector collecting and counting photons from both apertures. This is the way to implement experimentally the partial trace in the qubit part on the qutrit-qubit density operator. The purpose of the bucket detector is to warrant that we are observing a single photon qutrit in each coincidence registered between the bucket detector and the qutrit fringes resolving detector. In other words, it warrants that we are working with a quantum source of light. Besides that, with the partial trace operation on the qubit system we manage to prepare the qutrit state with the symmetries shown in the density operators presented in Eq. (\ref{states}). A lens $L_1$ projects the optical Fourier transform of the triple slit plane to the plane at $z_{d}$, where we scan an avalanche photodiode ($D_1$) along the $x$-direction. $810$ nm interference filters with bandwidth of $10$ nm are used to select the photon pairs. 100 $\mu$m single slit is coupled to the detector $D1$ such that the spatial fringes can be resolved. We record photons in the qutrit state in coincidence with those photons in the qubit state. The bucket detector assures us we are, in fact, measuring quantum states and realizing a partial trace operation.

\begin{figure}[h!]
\centering
\includegraphics[height=7.8cm]{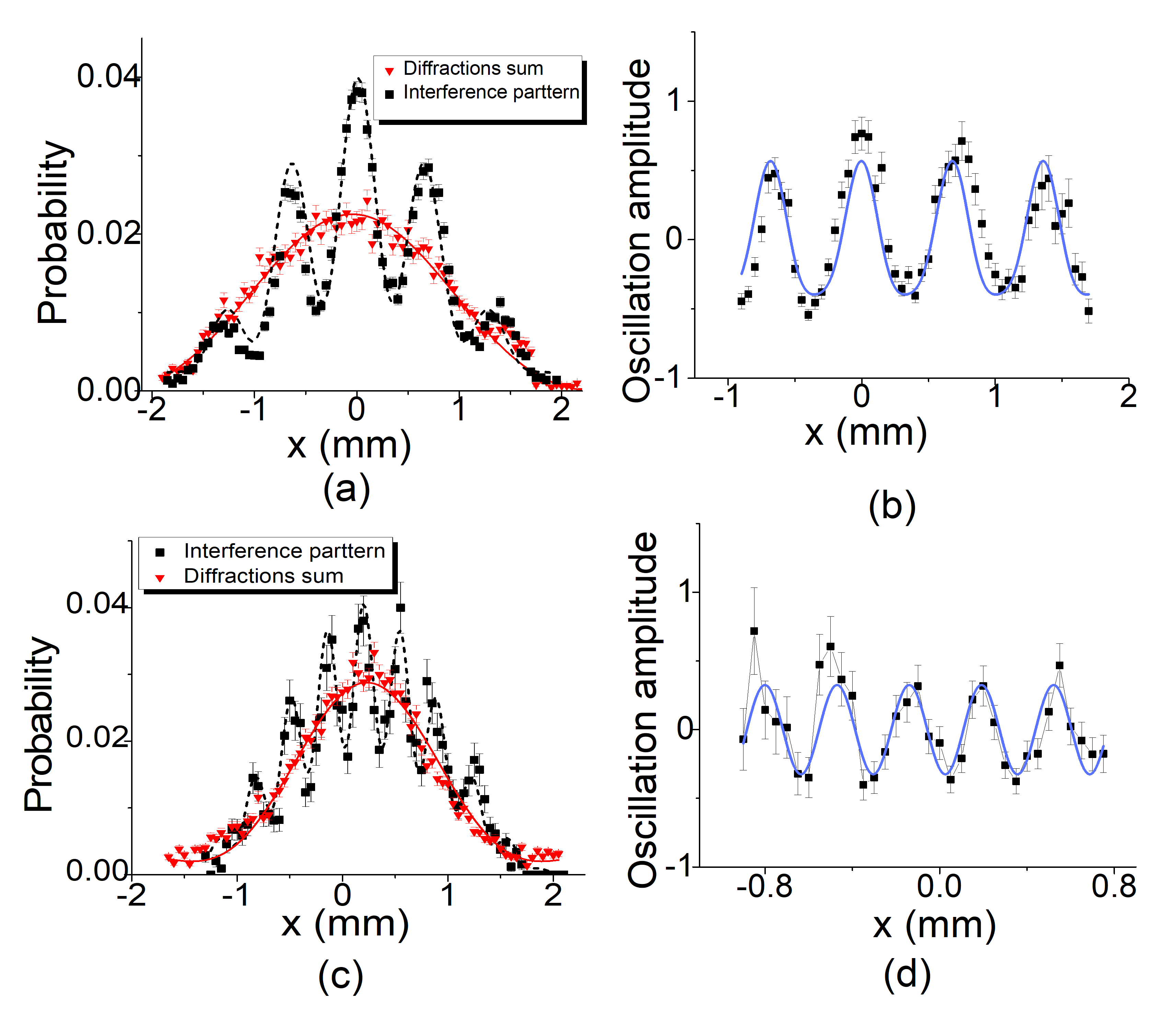}
\caption[Dados experimentais]{{\footnotesize Experimental results of interference and diffraction measurements. In Figs. \ref{fig:results}(a) and \ref{fig:results}(c) the black squares are the interference patterns from the states $\rho_1$ and $\rho_2$, respectively and the red circles are the respective Gaussian curves resulting from the sum of individual apertures diffraction patterns. Figs. \ref{fig:results}(b) and \ref{fig:results}(d) show the oscillation curves for the states $\rho_1$ and $\rho_2$ (Eq.(\ref{states})), respectively and are obtained from the experimental data shown in Figs. \ref{fig:results}(a) and \ref{fig:results}(c). The continuous curves superposing the experimental data, are fits of the theoretical expressions for the interference patterns and oscillations.}}\label{fig:results}
\end{figure}

The probability distributions $P\left(x,z_D\right)$ for the states in Eq. (\ref{states}), measured on the far field plane, are the black squares shown in Figs. \ref{fig:results}(a) and (c). The full lines superimposing  the experimental data (black curves) are the fits of the theoretical expressions to the experimental results, and the absolute values of the off-diagonal elements of $\rho_1$ and $\rho_2$ are the free parameters.  

\begin{figure}[h!]
\centering
\includegraphics[height=5.5cm]{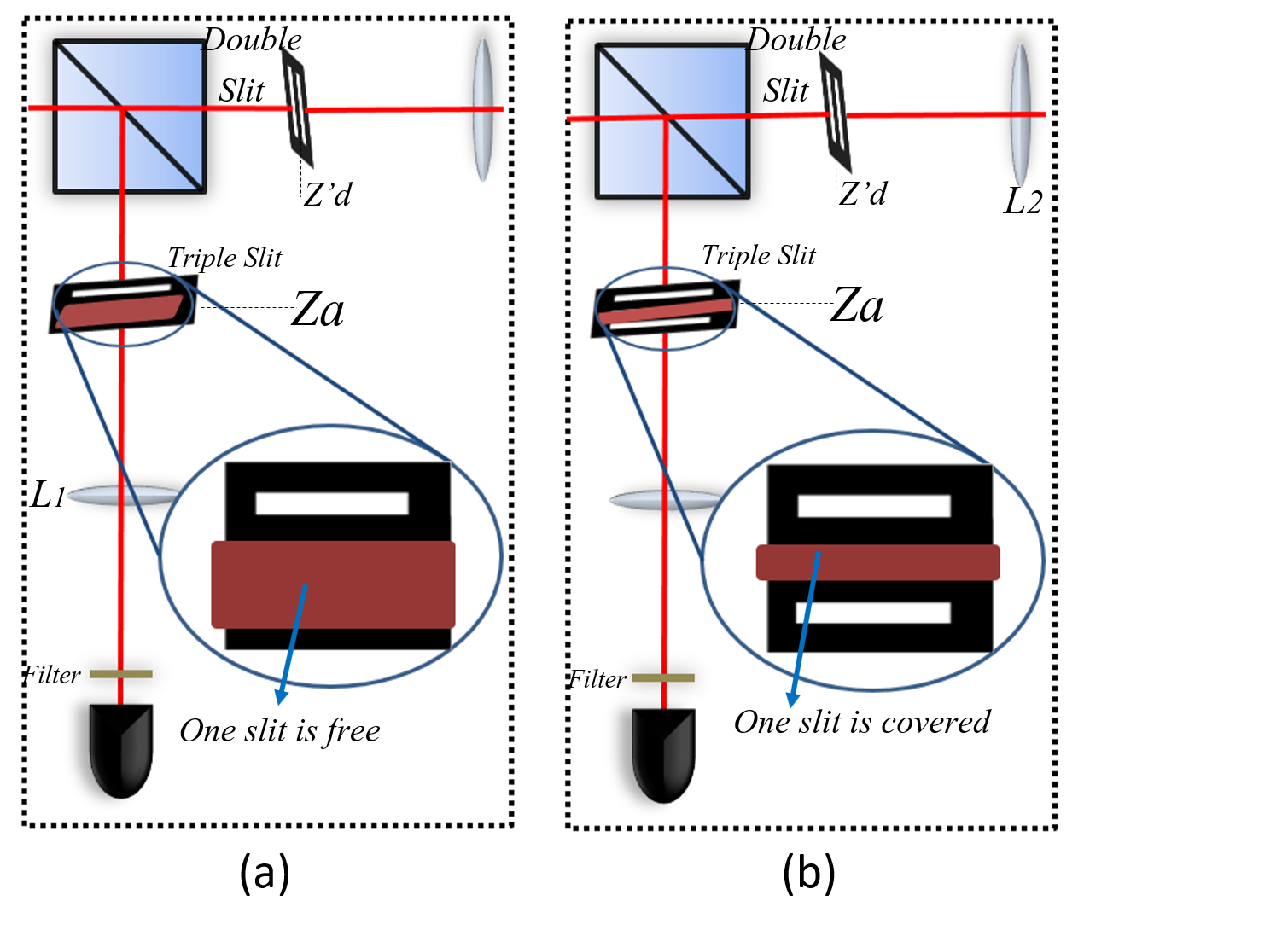} 
\caption[Setup dos envoltorios e medicoes dois a dois]{{\footnotesize Setup to measure the diffraction curves and the slits pair interference patterns. In \ref{fig:setupcoe2}(a), we block two of the three apertures and scan the detector in the optical Fourier transform plane to measure the diffraction wave. We repeat this process for each aperture. In \ref{fig:setupcoe2}(b), we block just one of the three apertures and measure the resulting interference pattern scanning the detector in the optical Fourier transform plane. This process must be made for all pair combinations of the apertures. Measurements in \ref{fig:setupcoe2}(a) and \ref{fig:setupcoe2}(b)  are done in coincidence with detector $D_2$.}}\label{fig:setupcoe2}
\end{figure}

To simulate the presence of an auxiliary incoherent state present in the proposal made in Ref.\cite{2017TaniaPaul}, we obtain $P_{diag}\left(x,z_D\right)$ by measuring the spatial photon counts distribution coming from each aperture individually in coincidence with the detector $D_2$ and adding them, \textit{i.e.}, we add the diffraction patterns of each aperture, which are measured separately. It reproduces the result that would be achieved for a diagonal state but without the requirement of other light source or other setup preparation. Although it is necessary  realize extra diffraction measurements, concerning to the the amount of optical elements in the setup, the experimental complexity is smaller. The experimental procedure that we used for measuring the diffraction patterns of the individual slits is shown in Fig.\ref{fig:setupcoe2}(a), and the normalized spatial photons distribution measured in coincidence with $D_2$, resulting from the diffraction sums, are the red dots presented in Figs. \ref{fig:results}(a) and (c), and the fits of the theoretical curves to the experimental diffraction sums are superposed to them (full red line).  

By obtaining $\Theta\left(x,z_d\right)$ with the experimental $P\left(x,z_D\right)$ and $P_{diag}\left(x,z_D\right)$, as shown in the first line of Eq. (\ref{oscfuntion}), we obtain the experimental oscillations shown in Figs. \ref{fig:results}(b) and \ref{fig:results}(d). The full line (blue) superposed to the experimental data are the theoretical oscillations curves used to fit the experimental data and $\mathcal{A}_1$ and $\mathcal{A}_2$ (Eq. (\ref{osccadaum})) are the free parameters. 

Table \ref{tab.coerencia} presents the values of quantum coherence that we obtained from the fits of the interference patterns and from the oscillation curves. We observe a good agreement between the values that the theoretical curve fitting provided for the sum of the modulus of the off-diagonal elements of $\rho$ in Eq. (\ref{sec2padrao}) and Eq. (\ref{osccadaum}) considering the states shown Eq. (\ref{states}). The sum of these terms are equal to the quantum coherence of the states.  

In general, fitting the experimental data with a mathematical function having adjustable parameters demands a knowledge about the theoretical expression of the patterns to be measured. Because of that, we need a different analysis when ``all coefficients phases equal to zero'' is the only state information at hand. Tab. \ref{tab.coerencia2} presents the values of the quantum coherence that we obtained by calculating the usual interference pattern visibility, the average of the oscillation curve maximum and the expression proposed in Ref.\cite{2017TaniaPaul}. We observe an agreement between the obtained experimental values and highlight that the experimental error is larger in the method suggested in Ref. \cite{2017TaniaPaul}.

We can measure quantum coherence from the usual interference pattern visibility for all states which produce interference patterns with only one oscillation frequency, independently of the dimension. For the cases where the null coefficients phases is the only information available about the state, the peaks average of the oscillation curve can also be employed, what can be verified by the agreement between the values in Table \ref{tab.coerencia2}. 

\begin{table}[htb]
\caption{{\footnotesize Experimental values of quantum coherence of the states shown in Eq. \ref{states}.}}\label{tab.coerencia}
\begin{tabular}{c  c  c  c}
\toprule \hline \hline \\
\small{State} & \ \ \ \ \ \ \small{Interference}  & \ \ \ \ \ \ \small{Oscillation}  \\
 & \ \ \small{fit} & \ \ \ \ \ \ \small{fit} & \\ \hline \\
$\rho_I$ & $0.579\pm 0.034$ & $0.568\pm 0.039$ \\\\
$\rho_{II}$ & $0.355\pm 0.025$ & $0.325\pm 0.035$ \\\hline\hline
\bottomrule
\end{tabular}
\end{table}

\begin{table}[htb]
\caption{{\footnotesize Experimental values of quantum coherence of the states shown in Eq. \ref{states}.}}\label{tab.coerencia2}
\begin{tabular}{c  c  c  c}
\toprule \hline \hline \\
\small{State} & \ \ \ \ $\underline{P^{max}-P^{min}}$  & \ \ \ \ \ \small{Peaks} & \ \ \ \ \ $\underline{P_{diag}^{max}-P_{diag}^{max}}$ \\
 & \ \ \ \ $P^{max}+P^{min}$& \ \ \ \ \ \small{average} & \ \ \ \ $P_{diag}^{max}$ \\\hline \\
$\rho_I$ & $0.555\pm 0.024$ & $0.530\pm 0.034$ & $0.565 \pm 0.091$ \\\\
$\rho_{II}$ & $0.353\pm 0.001$ & $0.346\pm 0.049$ & $0.313\pm 0.146$ \\\hline\hline
\bottomrule
\end{tabular}
\end{table}

\section{An alternative for a non-null phases} \label{alternative}

Now we present a general method that one can use to determine the quantum coherence of spatial states prepared with multiple slits with oscillations as the ones shown in Figs. \ref{fig:oscmatematica}(e) and (f). This method also enables us to obtain the absolute value of each off-diagonal coefficient of any spatial state.

In Eq. (\ref{sec2padrao}), each absolute value of the upper off-diagonal elements of $\rho$ ($\rho_{ij},\quad i<j$) controls the amplitudes of one of the cosine terms. We relate each cosine term to an impossibility to distinguish photons coming from one of the two apertures of a possible slit pair belonging to a multiple-slit. We can say that the interference in Eq. (\ref{sec2padrao}) results from the sum of interference patterns generated from pairs of apertures, or in other words, Eq.(\ref{sec2padrao}) is a combination of different double slit interferences patterns. Therefore, by thinking in quantum coherence, in Eq. (\ref{coherence}), as the sum of ``coherences" of the state pairs belonging to the state basis, it is intuitive to determine the total coherence by summing these ``partial coherences".     

Considering the state in Eq. (\ref{pho3}), if we use a procedure to prevent photons from passing through aperture ``$-1$", for example, the state after the multiple slits and available for detection at $z_d$ is
\begin{eqnarray}\label{pho4}
\rho_{III}=\frac{1}{\rho_{{\scriptscriptstyle-1-1}}+\rho_{{\scriptscriptstyle 00}}}\begin{pmatrix} \rho_{{\scriptscriptstyle-1-1}} & \left|\rho_{{{\scriptscriptstyle -10}}}\right|e^{i\varphi_{{\scriptscriptstyle-10}}} & 0 \\ 
\left|\rho_{{\scriptscriptstyle-10}}\right|e^{-i\varphi_{{\scriptscriptstyle-10}}} & \rho_{{\scriptscriptstyle 00}} & 0 \\ 
0 & 0 & 0\end{pmatrix}.
\end{eqnarray}   

The $\rho_{III}$ probability distribution for detecting photons in the far-field plane is
\begin{eqnarray}\label{osc-10}
P_{-10}(x,z_D)
 &\propto&\textit{sinc}^2\left(\frac{kax}{f_1}\right)\left[1+V_{-10}\cos\left(\gamma x-\varphi_{-10}\right)\right],
\end{eqnarray}
where $V_{-10}=2\left|\rho_{{\scriptscriptstyle -10}}\right|/\left(\rho_{{\scriptscriptstyle -1-1}}+\rho_{{\scriptscriptstyle 00}}\right)$.

Eq. (\ref{pho4}) is equivalent to a density operator expected for a two-dimensional spatial state, whose quantum coherence is determined by the visibility $V_{-10}$, \textit{i}. \textit{e}., the oscillation contrast presented in Eq.(\ref{osc-10}) \cite{1991Mandel}. Therefore, we can determine the quantum coherence $\mathcal{C}$ of the state in Eq. (\ref{pho3}) by summing the visibilities of interference patterns of all the possible aperture pairs that constitute the multiple-slit. For a three-dimensional state, as in Eq. (\ref{pho3}), we have three aperture pairs and
\begin{equation}\label{general}
\mathcal{C} = \frac{1}{2}\frac{V_{-10}\left(\rho_{-1-1}+\rho_{00}\right)+V_{-11}\left(\rho_{-1-1}+\rho_{11}\right)+V_{01}\left(\rho_{00}+\rho_{11}\right)}{\rho_{-1-1}+\rho_{00}+\rho_{11}},	
\end{equation} 
where the diagonal coefficients $\rho_{-1-1}, \rho_{00}$ and $\rho_{11}$ can be determined by performing a triple slit image measurement in coincidence with detector $D_2$, as explained above. $V_{ij}$ is the interference pattern visibility resulted from the interference between the apertures $i$ and $j$ recorded by detector $D_1$ in coincidence with detector $D_2$.

For a state with dimension $D$, it is necessary to measure $N$ interference patterns, being
\begin{eqnarray}
N=\frac{D!}{2\left(D-2\right)!}.
\end{eqnarray}
Although $N$ increases quickly with the dimension of the state, this method provides us an accurate result for the cases in which Eq. (\ref{oscmaxemin}) is not applicable and does not require the presence of any additional equipment to make the density matrix element phases equal, by using a spatial light modulator (SLM) for example. From the measurements of slit pairs interference patterns, we can also determine the absolute value of each off-diagonal coefficient of $\rho$. Besides that, the displacement between two of these interference patterns give us the relative phase between two matrix elements. The displacement between the maximum points of $P_{-10}(x,z_D)$ and $P_{10}(x,z_D)$, for example, allow us to obtain $\left|\varphi_{{\scriptscriptstyle-10}}-\varphi_{{\scriptscriptstyle 10}}\right|$. 

Eq. \ref{general} allows us also to evaluate the expected quantum coherence for the prepared qutrit quantum states and the maximum expected quantum coherence for these states can be obtained from it. States whose density matrices have the symmetries of  Eq. (\ref{states}) do not have quantum coherence equal to one and the oscillation amplitudes measured experimentally reflect the quantum coherence of our prepared states. Notice that for the state $\rho_I$, $V_{-1,1} = 0 $ ideally, so that quantum coherence for this state is 
\begin{eqnarray}
\mathcal{C}_I = \frac{1}{2}\frac{V_{-10}\left(\rho_{-1-1}+\rho_{00}\right)+V_{01}\left(\rho_{00}+\rho_{11}\right)}{\rho_{-1-1}+\rho_{00}+\rho_{11}},
\end{eqnarray}
By state normalization $\rho_{-1-1}+\rho_{00}+\rho_{11}=1$, such that $
\mathcal{C}_I = V_{-10}\frac{\left(\rho_{-1-1}+\rho_{00}\right)}{2}+V_{01}\frac{\left(\rho_{00}+\rho_{11}\right)}{2}$.
Even with maximal coherence between the modes $\ket{-1}$ and $\ket{0}$ and between $\ket{1}$ and $\ket{0}$, i. e., even for $V_{-10}=V_{01}=1$ the theoretical quantum coherence $\mathcal{C}_I$ is smaller than $1$, because $
\left(\rho_{-1-1}+\rho_{00}\right)<1\quad \mbox{and} \quad \left(\rho_{00}+\rho_{11}\right)<1$.
Some complementary equations, as discussed in \cite{1996Englert} and \cite{2015-Querish}, show that the distinguishability between the possible states of a system decreases the amount of the state quantum coherence. In other words, for any dimension the upper bound of the quantum coherence of a state with some symmetry is reached when all the diagonal elements of state density matrix are equal, as we can conclude from Eq. (3) of Reference \cite{Bagan}. Considering a state with the symmetry of $\rho_I$ and with all diagonal elements equal to $1/3$, we have $\mathcal{C}_I = \frac{V_{-10}+V_{01}}{3}$.
Therefore, the upper bound for quantum coherence in this case is $2/3\approx 0.67$, and the experimental values shown in the Tables I and II for $\rho_{I}$ is slightly smaller than the upper bound. 
Similarly, for the state $\rho_{II}$,  $V_{-10}=V_{01}=0$ and $
\mathcal{C}_{II} = V_{-11}\frac{\left(\rho_{-1-1}+\rho_{11}\right)}{2}$.
If all diagonal elements are equal to $1/3$, $
\mathcal{C}_{II} = \frac{V_{-11}}{3}	
$
and the upper bound for quantum coherence is $1/3\approx 0.33$. We observe that the measured quantum coherence of the state $\rho_{II}$ is near the upper bound within the experimental error range. This happens because there are influence of matrix density elements that are not absolutely null in the experimental prepared state. Therefore, the results presented in Tables I and II are in accordance with the expected upper bound for the quantum coherence of the states that we have prepared. 

In this article, we focus on studying the quantum coherence measurement of spatial photonic states prepared with multiple slits, so that the method as presented in this section is general because is valid for qudits of any dimension prepared with multiple slits. However, it could be used for photonic qudits in others degrees of freedom. For example, for photonic qutrits prepared in angular momentum modes, a spatial light modulator (SLM) can be programmed to filter out (or block) one of the angular momentum modes in order to make possible the measurement of the analogous remaining angular momenta  mode pair interference \cite{momentoangular}.

\section{Conclusion} \label{conclusion}

In this article, we measured experimentally the quantuom coherence of prepared qudit states.  Photons transmitted through multiple slits with $D$ apertures are prepared in a photon path qudit state having a quantum coherence associated with it. We determined the quantum coherence of qutrit spatial states experimentally, employing a method based on the theoretical proposal made in Ref.\cite{2017TaniaPaul}. We discussed the method validity and presented an alternative that is always valid for a general qudit state, where no assumption is made about the phases of the elements that constitute the density matrix that describes the qudit state. We also analyzed the multiple-slits experiment from the perspective of an interferometer containing path-markers and showed that the quantum coherence measurement must be realized in an optical configuration such that the interference patterns are detected with path-markers states maximally indistinguishable.

\section{Acknowledgments}
We thank Renan Souza Cunha. for enjoyable and fruitful discussions about quantum coherence. This work was supported by CNPq, CAPES, FAPEMIG, National Institute of Science and Technology in Quantum Information (Capes and CNPq, Brazil).

\appendix 
\section{Near- and far-field operations as path-marker in a context of spatial states} \label{apendice}
We show here how the Englert's discussion \cite{1996Englert} is present in the context of photonic spatial states (prepared with slits) and the requirements for the use of interference patterns for quantum coherence measurements. Consider the density operator shown above in Eq. (\ref{estadogeral}) that describes a general state of a quantum system and the Quantum coherence defined in Eq. (\ref{coherence})  for the state represented in Eq.(\ref{estadogeral}) \cite{2014Baumgratz,2015-Quresh}.

In the general treatment of the quantum coherence measurement, an ensemble of particles is prepared in a superposition path state after it crosses an interferometer input port and its state is represented by Eq. (\ref{estadogeral}). The system interacts with a path-marker inside the setup, in such a way that each possible state of the path-marker in some base state, marks a particle in one specific state of the system base state. The resulting joint state after the interaction is \cite{1996Englert,2001-Durr,2003-BimontePRA,2015-Quresh} 
\begin{eqnarray}\label{rhojoint}
\rho_{T} &=& \sum_{l,m=-\lambda}^{\lambda}\rho_{lm}\ket{l}\bra{m}\otimes\ket{d_l}\bra{d_m},
\end{eqnarray}
where $\left\{\ket{d_l}\right\}$ are the possible path-marker states. By tracing out $\rho_T$ over the path-marker states, we obtain the available state to the particles register, since we detect them without measuring the path-marker state directly. Thus, the probability distribution of detecting a particle at the interferometer output ports involves the matrix elements of the reduced state operator 
\begin{eqnarray}\label{rho2geral}
\rho_{2} &=& \sum_{l,m=-\lambda}^{\lambda}\rho_{lm}\ket{l}\bra{m}\braket{d_m|d_l},
\end{eqnarray}
which depends on the internal products between the possible path-marker states.

As a consequence, the observation of quantum coherence depends on the path-marker states and can be equal to or less than the associated with the system initial state $\rho$. If the states belonging to $\left\{\ket{d_l}\right\}$ are entirely distinguishable, that is, orthogonal, we observe a null quantum coherence even if the initial state has coherence. In other words, in this case we do not have access to the off-diagonal elements of $\rho$. Therefore, experimental determination of quantum coherence requires path-markers, or measurement apparatus, which states do not preclude the observation of the off-diagonal of $\rho$ or the direct quantum coherence value.

As mentioned above, concerning to the spatial photonic states, a common method to prepare them consists into discretizing the transverse profile of an ensemble of photons, identically generated, using multiple slits \cite{2005leo5, 2007leo2, 2008Japoneses, holandeses}. Fig. \ref{fig:discrete} shows a multiple slits array which selects photons within specific intervals of linear momentum. 
Fig. \ref{fig:discrete}.b and \ref{fig:discrete}.b show two possible optical configurations of detecting the photonic spatial probability distribution after the multiple slit
at the longitudinal position $z_d$ and along the transverse $x$-direction. In this scenario, the probability of detecting one photon on a detection plane is given by Eq. (\ref{Pgeral}).

For the state preparation, we use a multiple-slit set with thin apertures with $2a$ width separated by $d$, so we can consider only the detection of photons that crossed the multiple-slit set with $D$ apertures and passed through some aperture $l$. After photons are transmitted through slits their path states forms a discrete set. Each slit $l$ defines a state vector of the base $\left\{\ket{l}\right\}$ and $\ket{l}$ represents a photon that crossed the slit $l$. Therefore, the operator \textit{a}($q$) acts only on photons in some mode described by $\ket{l}$, which can be written in terms of the Fock state $\ket{1q}$ in the transverse momentum variable $q$: $\ket{l} = \sqrt{\frac{\pi}{a}}\int dq e^{-\i qld} sinc (qa) \ket{1q} $  \cite{1995Mandel,2007leo2}.
By rewriting $\Gamma=E^{\left(\scriptstyle{-}\right)}\left({\scriptstyle x,z_d}\right)E^{\left({\scriptstyle +}\right)}\left({\scriptstyle x,z_d}\right)$ in the $\left\{\ket{l}\right\}$ base, we obtain \cite{2000sebastiao1,miguel2010,leo04,2007leo2}
\begin{eqnarray}
\Gamma&=&\sum_{m=-\lambda}^{\lambda}\sum_{l=-\lambda}^{\lambda}\left(\bra{m}E^{(-)}(x,z_d)E^{(+)}(x,z_d)\ket{l}\right)\ket{m}\bra{l}. 
\end{eqnarray} 
We can write the probability shown in Eq. (\ref{Pgeral}) in the form
\begin{eqnarray}\label{probjunto}
P&=&\sum_{l,m=-\lambda}^{\lambda}\bra{m}E^{(-)}(x,z_d)E^{(+)}(x,z_d)\ket{l}\rho_{lm},
\end{eqnarray}
which is a sum of different terms. Notice that it is possible to write a state operator $\rho^{'}_{2}$ in such a way that its matrix elements are the different terms in the sum of Eq. (\ref{probjunto})
\begin{eqnarray}\label{rho2linha}
\rho^{'}_{2} &=& \sum_{l,m=-\lambda}^{\lambda}\rho_{lm}\ket{l}\bra{m}\bra{m}E^{(-)}(x,z_d)E^{(+)}(x,z_d)\ket{l}.
\end{eqnarray}

We can also see $\rho^{'}_{2}$ as the reduced density operator obtained from the joint state 
\begin{eqnarray}\label{rhojointlinha}
\rho^{'}_{T}&=&\sum_{l,m=-\lambda}^{\lambda}\rho_{lm}\ket{l}\bra{m}\otimes E^{(+)}(x,z_d)\ket{l}\bra{m}E^{(-)}(x,z_d),
\end{eqnarray}
after we apply the trace operation over the subspace defined by the set $\left\{E^{(+)}(x,z_d)\ket{l}\right\}$.

It is notable the similarity between the Eqs. (\ref{rhojointlinha}) and (\ref{rhojoint}), and between the Eqs. (\ref{rho2linha}) and (\ref{rho2geral}). It is also remarkable the fact that if the elements of $\left\{E^{(+)}(x,z_d)\ket{l}\right\}$ are orthogonal to each other, we do not observe any coherence, since we do not have access to the off-diagonal elements of $\rho^{'}_{2}$ (Eq. (\ref{rho2linha})) in this case.
On the other hand, if the elements of $\left\{E^{(+)}(x,z_d)\ket{l}\right\}$ are nonorthogonal, have equal amplitudes and different phases, we can observe an oscillation controlled only by the matrix elements of the initial system state (Eq. (\ref{estadogeral})).

In Eq. (\ref{rhojointlinha}) we interpret the spatial photonic state as interacting with a path-marker before the detection at $z_d$, similarly to Eq. (\ref{rhojoint}). We can identify each vector state $\ket{d_l}$ as the state of a photon which is annihilated at the detection plane after it propagates from the aperture $l$ and passes through an optical arrangement. Therefore, the optical setup configuration between the multiple slits plane, at $z_a$, and the detection plane, at $z_d$, determines the possible path-marker states, that is, $\ket{d_l}=E^{(+)}(x,z_d)\ket{l}$.

We set the states $\left\{E^{(+)}(x,z_d)\ket{l}\right\}$ orthogonal when we use a lens at position $z_l$, with focal length equal to $\left(z_l-z_a\right)/2 = \left(z_d-z_l\right)/2$, to project the multiple slit image on the plane at $z_d$ (Fig. \ref{fig:discrete}.b), resulting in
\begin{eqnarray}\label{Eimage}
 \sum_{l=-\lambda}^{\lambda}E^{(+)}(x,z_d)\ket{l} &\propto& \sum_{l=-\lambda}^{\lambda}\prod\left(\frac{x+ld}{2a}\right)\ket{0},
\end{eqnarray}
where $\ket{0}$ indicates no photons in any mode of the set $\left\{\ket{l}\right\}$ and is resulting from the operation of \textit{a}$\left(l\right)$ over $\ket{l}$; $d$ and $a$ are, respectively, the separation between two adjacent identical apertures and their width; $\prod\left(\frac{x+ld}{2a}\right)$ is the function that describes the rectangular aperture centered in $x=ld$, with width $2a$ \cite{2005leo5}.

The probability distribution to detect a photon at $x$ transverse position (Fig. \ref{fig:discrete}.b)
\begin{eqnarray}
P(x,z_D)&\propto&\sum_{l,m=-\lambda}^{\lambda}\prod\left(\frac{x+ld}{2a}\right)\prod\left(\frac{x+md}{2a}\right)\rho_{lm}.
\end{eqnarray}

By construction, rectangular functions in Eq. (\ref{Eimage}) are orthogonal to each other within the range $2a$ and consequently, $P(x,z_d)\neq 0$ only if $l=m$. In this case, we can identify which aperture each counted photon comes from and determine the initial state diagonal elements $\left\{\rho_{ll}\right\}$. On the other hand, if we use a lens, with focal length equal to $\left(z_l-z_a\right)=\left(z_d-z_l\right)$, to project the multiple slits to the Fourier plane at $z_d$ (focal plane), we set the states $\left\{E^{(+)}(x,z_d)\ket{l}\right\}$ in a non-orthogonal form, in such a way that we cannot distinguish photons coming from a specific aperture (Fig. \ref{fig:discrete}.c). In this case \cite{miguel2010,leo20072},
\begin{eqnarray}
\sum_lE^{(+)}(x,z_d)\ket{l}&\propto&\sum_le^{i\gamma xl}\ket{0},
\end{eqnarray}
and $P(x,z_D)$ is given Eq. (\ref{sec2padrao}).
Eq.(\ref{sec2padrao}) describes mathematically an interference pattern at $z_d$ with contributions of both, diagonal ($\rho_{ll}$) and off-diagonal ($\rho_{lm}, \quad l\neq m$) coefficients of $\rho$. However, quantum coherence, which we intend to determine, involves only the absolute values of the off-diagonal coefficients (Eq. (\ref{coherence})). Therefore we can infer quantum coherence from an interference pattern, when the possible path-marker states are non-orthogonal and maximally indistiguishable. The path-marker states in photonic spatial interference patterns are produced ny the optical configuration that produce the interference pattern as demonstrated here.

\frenchspacing 

\bibliographystyle{plain}

\end{document}